\documentclass[twocolumn,prb, showpacs]{revtex4-1}
\usepackage{graphicx}
\usepackage{dcolumn}
\usepackage{bm}
\usepackage{color}

\begin{document}


\title{Dynamic surface electronic reconstruction as symmetry-protected topological orders in topological insulator $Bi_2Se_3$ }

\author{G. J. Shu$^{1,2,3,6}$}
\author{S. C. Liou$^4$}
\author{S. K. Karna$^{1}$}
\author{R. Sankar$^1$}
\author{M. Hayashi$^1$}
\author{F. C. Chou$^{1,5,6}$}
\email{fcchou@ntu.edu.tw}
\affiliation{
$^1$Center for Condensed Matter Sciences, National Taiwan University, Taipei 10617, Taiwan}
\affiliation{
$^2$Department of Materials and Mineral Resources Engineering, National Taipei University of Technology, Taipei 10608, Taiwan}
\affiliation{
$^3$Institute of Mineral Resources Engineering, National Taipei University of Technology, Taipei 10608, Taiwan}
\affiliation{
$^4$AIM Lab, Nano Center, University of Maryland, College Park, MD 20742, U.S.A. }
\affiliation{
$^5$National Synchrotron Radiation Research Center, Hsinchu 30076, Taiwan}
\affiliation{
$^6$Taiwan Consortium of Emergent Crystalline Materials, Ministry of Science and Technology, Taipei 10622, Taiwan}

\date{\today}

\begin{abstract}
Layered narrow band gap semiconductor Bi$_2$Se$_3$ is composed of heavy elements with strong spin-orbital coupling (SOC), which has been identified both as a good candidate of thermoelectric material of high thermoelectric figure-of-merit ($ZT$) and a topological insulator of \textit{Z$_2$}-type with a gapless surface band in Dirac cone shape.  The existence of a conjugated $\pi$-bond system on the surface of each Bi$_2$Se$_3$ quintuple layer is proposed based on an extended valence bond model having valence electrons distributed in the hybridized orbitals.  Supporting experimental evidences of a 2D conjugated $\pi$-bond system on each quintuple layer of Bi$_2$Se$_3$ are provided by electron energy-loss spectroscopy (EELS) and electron density (ED) mapping through inverse Fourier transform of X-ray diffraction data.  Quantum chemistry calculations support the $\pi$-bond existence between partially filled 4$p_z$ orbitals of Se via side-to-side orbital overlap positively. The conjugated $\pi$-bond system on the surface of each quintuple Bi$_2$Se$_3$ layer is proposed being similar to that found in graphite (graphene) and responsible for the unique 2D conduction mechanism.  The van der Waals (vdW) attractive force between quintuple layers is interpreted being coming from the anti-ferroelectrically ordered effective electric dipoles which are constructed with $\pi$-bond trimer pairs on Se-layers across the vdW gap of minimized Coulomb repulsion. 

\end{abstract}

\pacs{71.15.Ap; 87.59.-e; 73.20.-r; 79.20.Uv}


\maketitle

\section{\label{sec:level1} introduction\protect\\ }


Bi$_2$Se$_3$ and Bi$_2$Te$_3$ are layered materials consisting of closely packed quintuple layers of Bi-Se(Te) with van der Waals (vdW) gaps in between.\cite{Huang2012}  Both compounds are composed of heavy elements of expected strong spin-orbit coupling (SOC) and have been explored intensely as a good candidate of material with high thermoelectric figure-of-merit $ZT$.\cite{Dresselhaus2007}  Bi$_2$Se$_3$ has also been classified as a typical $Z_2$-type topological insulator,\cite{Zhang2009} i.e., an insulator (semiconductor) with time reversal symmetry-protected topological orders and a unique mechanism of surface conduction.  Close relationships between thermoelectric material and topological insulator has been identified repeatedly since.\cite{Muchler2012}

The discovery of topological insulators has been one of the most important breakthroughs in condensed matter physics over the past decade. Similar to graphene as an ideal 2D material of monoatomic thickness,\cite{Neto2009} Kane and Mele first suggested that a Dirac-cone-shaped surface band can also be found in this novel class of material.\cite{Kane2005}  Many topological insulators have been predicted through theoretical calculations and subsequently verified experimentally by the angle resolved photoemission spectroscopy (ARPES) leading to the 2016 Nobel prize in physics.\cite{Hasan2010, Nobel2016}   However, the fundamental physical properties of topological insulators have been explored mainly in the energy-momentum space theoretically and experimentally, a real space view to understand the meaning of a topological phase and its unique surface electron configuration is highly desirable, especially with the purpose on designing more topological insulators suitable for device application through material engineering from the chemical aspect.


We find that a real space view of chemical bonding from the hybridized molecular orbital perspective is helpful to explain the puzzling nature of surface conduction for an insulator having a  symmetry-protected topological phase.  Here, by examining the real space bonding electron distribution of Bi$_2$Se$_3$ through the inverse Fourier transform of X-ray diffraction data, we are able to interpret the electron density (ED) mapping of Bi$_2$Se$_3$ with an extended valence bond model containing both $\sigma$- and $\pi$-bonds reasonably, which allows the electron counts in the chemical bond consistently from both the chemical and physical perspectives.  

The existence of $\sigma$- and $\pi$-bonds is supported by the electron energy-loss spectroscopy (EELS) via a reasonable electron counts on bonding electrons in collective resonance.  Surface $\pi$-bond is proposed to be formed among Se atoms on both sides of each quintuple layer with instantaneous side-to-side orbital overlap of half-filled lobes of hybridized $sp^3d^2$ orbital. When many quintuple layers of Bi$_2$Se$_3$ are stacked into a bulk, the conjugated $\pi$-bonds could form effective electric dipoles across the vdW gap with anti-ferroelectrically ordered $\pi$-bond pairs under quantum fluctuation, which is the origin of vdW force.

\section{\label{sec:level1} Extended valence bond model of $Bi_2Se_3$\protect\\}

\begin{figure}
\begin{center}
\includegraphics[width=3.5in]{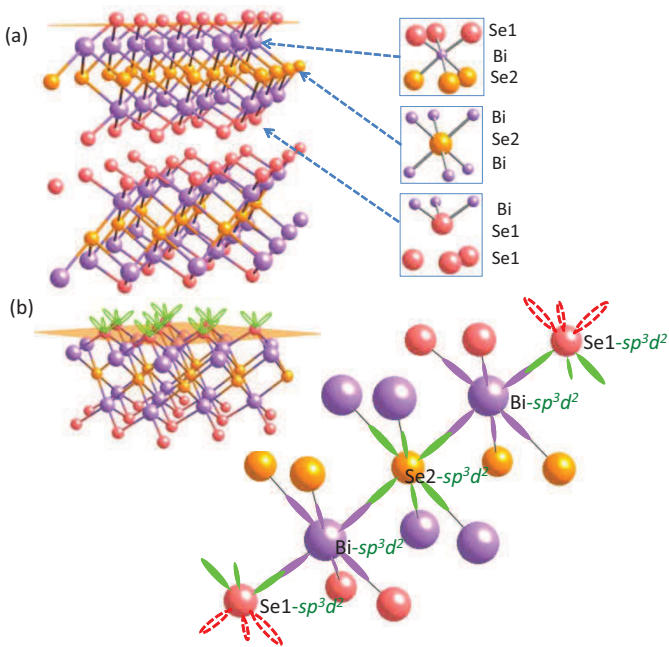}
\end{center}
\caption{\label{fig-structure}(color online) (a) The crystal structure of Bi$_2$Se$_3$ is shown with two quintuple layers, and the octahedral coordination of each atom is shown on the right. (b) The octahedral coordination for all Bi and Se atoms form valence bonds through six valence shell electrons in $s$-$p$-$d$ orbitals and hybridize into $sp^3d^2$ orbital of six directional half-filled lobes. While all Bi-Se atoms are able to form covalent $\sigma$-bonds via the overlap of two half-filled $sp^3d^2$ lobes in the bulk, the Se1-layer must have three out of the six $sp^3d^2$ lobes remain half-filled facing the vdW gap or on the crystal surface.  }
\end{figure}

The crystal structure of Bi$_2$Se$_3$ can be described as Bi and Se atoms packed in  ABCABC stacking as cubic close packing (ccp), in contrast to the ABAB stacking of hexagonal close packing (hcp), which has a space group $R\bar{3}m$ containing vdW gap between quintuple layers, as shown in Fig.~\ref{fig-structure}.\cite{Huang2012}  Since both Bi ([Xe]4f$^{14}$5d$^{10}$6s$^2$6p$^3$6d$^0$) and Se ([Ar]3d$^{10}$4s$^2$4p$^4$4d$^0$) atoms are packed in octahedral coordination, we propose that each quintuple unit of Bi$_2$Se$_3$ can be described fully by the hybridization of $s$-$p$-$d$ orbitals ($n$=6 for Bi and $n$=4 for Se) to preserve the octahedral coordination having six required directional $\sigma$-bonds between Bi-Se, which follows the standard valence shell electron pair repulsion (VSEPR) rule and has been discussed fully from the chemical bond perspective.\cite{inorganic} 

The concept of orbital hybridization has been applied to the silicon crystal (Si=[Ne]3s$^2$3p$^2$) as a textbook example, where covalent $\sigma$-bonds between Si-Si are formed via the overlap of tetrahedral-shaped half-filled $sp^3$ hybridized orbital. However, for the fixed number of valence shell electrons in Bi  and Se, and the resolved octahedral coordination from structure analysis, there are many possible ways to arrange how chemical bonds are formed, as long as the total number of valence electrons and coordinations are conserved.  The typical hybridization of $s$-$p$-$d$ electrons to satisfy an octahedral coordination is the hybridized $sp^3d^2$ orbitals having six directional lobes in octahedral shape, various orbital hybridization models have been proposed for the Bi$_2$Se$_3$-type quintuple layer compounds since 1958.\cite{Mooser1958, Drabble1958, Bhide1971} Unfortunately all early proposed molecular models were correct on the valence electron counts without considering the unique requirement of surface conduction, mostly because the latter condition was only introduced in the latest development of topological material research.\cite{Hasan2010, Nobel2016}

Based on the valence shell electrons of Bi and Se, the octahedral coordination of BiSe$_6$ and SeBi$_6$, and the $sp^3d^2$ orbital hybridization of six directional lobes, we propose an extended valence bond model for the Bi$_2$Se$_3$ quintuple unit as shown in Fig.~\ref{fig-structure}(b). A correct molecular model must contain bonding characters which are consistent to the  unique physical properties of a condensed matter state, i.e., the inferred electron distribution of Bi$_2$Se$_3$ must reflect the unique requirement of conducting surface and insulating bulk simultaneously. The main characters of Bi include high $Z$ number, large atomic size, and the large principal quantum number that implies close spacing between energy levels ($E_n$ $\propto n^2$), i.e., strong spin-orbit coupling is more likely to occur across a narrow gap, which allows spontaneous exchange between the Zeeman-like magnetic energy gain/loss and the electric potential rise/drop.    

To explain the insulator nature of the bulk part of a topological insulator, we find the insulating nature for the bulk of Bi$_2$Se$_3$ is consistent to the assignment of $\sigma$-bond between Bi-Se, and electrons bound in the $\sigma$-bond region in real space correspond to the valence band within band picture description in the reciprocal space.  Comparing to the typical semiconductor of Si crystal ($E_g$$\sim$1 eV) which have covalent $\sigma$-bonds through overlapped half-filled hybrid $sp^3$ orbital of four tetrahedral directional lobes, it is reasonable to form similar $\sigma$-bonds in Bi$_2$Se$_3$ through the overlap of hybrid $sp^3d^2$ orbital having six half-filled directional octahedral lobes.  On the other hand, unlike the perfect covalent bond of Si crystal with much higher melting point of 1414$^\circ$C, the electronegativity (EN) difference between Bi (EN=2.02) and Se (EN=2.55) implies the $\sigma$-bond is weakly polarized.  In addition, the lower binding energy of high principal quantum number ($n$$\sim$ 4-6) for Bi(Se) is consistent to the characters of Bi$_2$Se$_3$, including a narrow band gap of $\sim$0.2-0.3 eV,\cite{Nechaev2013} a low melting point of 705$^\circ$C, and the longer bond length between Bi-Se.\cite{Huang2012}  

To explain the unique surface-only conduction of a topological insulator, surface band inversion as a result of strong spin-orbit coupling has been proposed for the partially filled $p$-orbitals in Bi$_2$Se$_3$.\cite{Zhang2009}  In the real space view, the Se atoms on both sides of each quintuple layer (Se1-layer shown in Fig.~\ref{fig-structure}) have three out of six lobes of hybrid $sp^3d^2$ orbital being unpaired and dangling in the excited state, unless a unique 2D chemical bonding emerges, which must be closely related to the surface conduction mechanism of excitonic nature in 2D.  While there are no itinerant electrons available according to present molecular model, only a system of partially localized electrons in the three unpaired $dp^3d^2$ lobes per Se1 being shared by six neighboring Se1 atoms on the surface (see Fig.~\ref{fig-structure}(b)), it is likely that partial localization could lead to surface conduction via local electron exchange in 2D.  

It is inspiring to find that a similar puzzling observation has been found in the 1D conducting polymer of unexpected high conductivity, which has been proposed coming from the partially localized electrons being shared dynamically as a conjugated $\pi$-bond system in 1D.\cite{Roncali2007}  In addition, similar surface conduction can also be identified in graphene through a conjugated $\pi$-bond system in 2D, where each half-filled $p_z$ orbital per carbon is shared by three neighboring carbon atoms in the honeycomb lattice to form a surface $\pi$-bond instantaneously.\cite{Carbone2009}  We propose that a similar $\pi$-bond system could also be identified on the surface of Bi$_2$Se$_3$, i.e., side-to-side overlap of the three half-filled hybridized $sp^3d^2$ lobes per Se1 may form three $\pi$-bonds among Se1-Se1 on the surface of each quintuple layer.  In addition, the three half-filled $sp^3d^2$ lobes per Se1 must be shared by six  neighboring Se1 atoms in the same plane without breaking the six-fold symmetry in 2D, which could be viewed as \underline{symmetry-protected topological orders in real space}.  The sharing requirement of the $\pi$-bond electrons can be viewed as a conjugated $\pi$-bond system responsible for the unique 2D conduction.  In search of the proposed conjugated $\pi$-bond system in Bi$_2$Se$_3$, electron energy-loss spectroscopy (EELS) and electron density (ED) mapping have been studied in the following.    


 

\section{\label{sec:level1} Electron density mapping of $Bi_2Se_3$\protect\\}


\begin{figure}
\begin{center}
\includegraphics[width=3.5in]{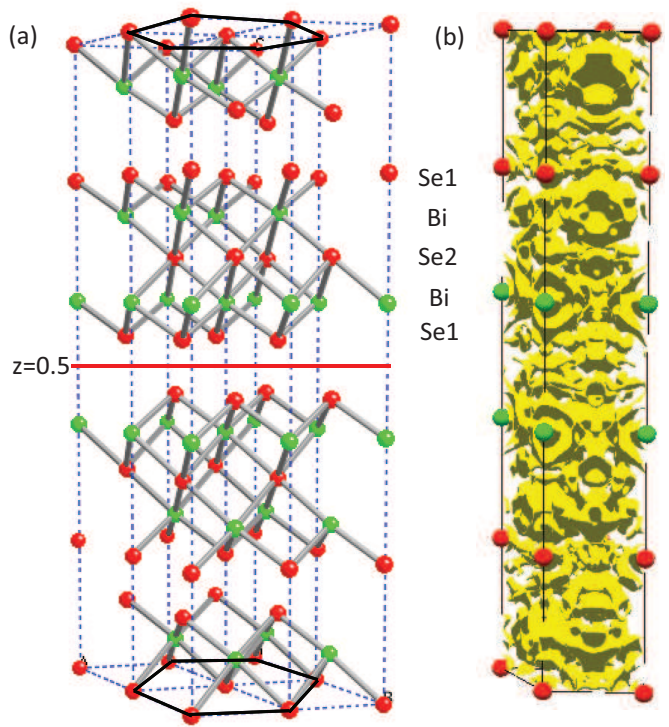}
\end{center}
\caption{\label{fig-EDin3D}(color online) (a) The crystal structure of Bi$_2$Se$_3$ is shown having a right rhombic prism unit cell of three formula units per cell in the hexagonal family (space group $R\bar{3}m$). (b) Selected atoms of Se1(red) and Bi(green) are added to the electron density information obtained from the inverse Fourier transform of the X-ray diffraction data.}
\end{figure}

\begin{figure}
\begin{center}
\includegraphics[width=3.5in]{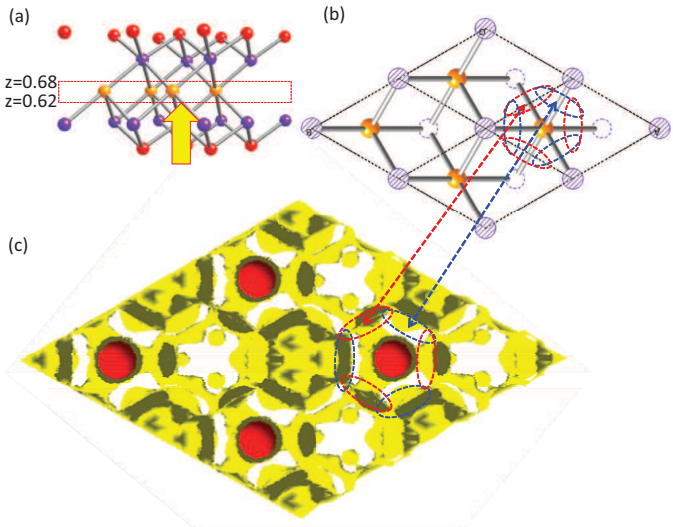}
\end{center}
\caption{\label{fig-Se2cut}(color online) The electron density contour map of Se2-layer, where (a) shows the side view of the ball and stick model of a quintuple layer cut between $z$=0.62 and 0.68, (b) shows the expected [001] projection of Se2 layer from below, with Bi atoms behind (dashed circles) and in front (empty circles) of the projection plane, and (c) is the corresponding integrated electron density contour map, where six $\sigma$-bonds between Se2-Bi atoms are indicated in blue and red ovals.   }
\end{figure}

\begin{figure}
\begin{center}
\includegraphics[width=3.5in]{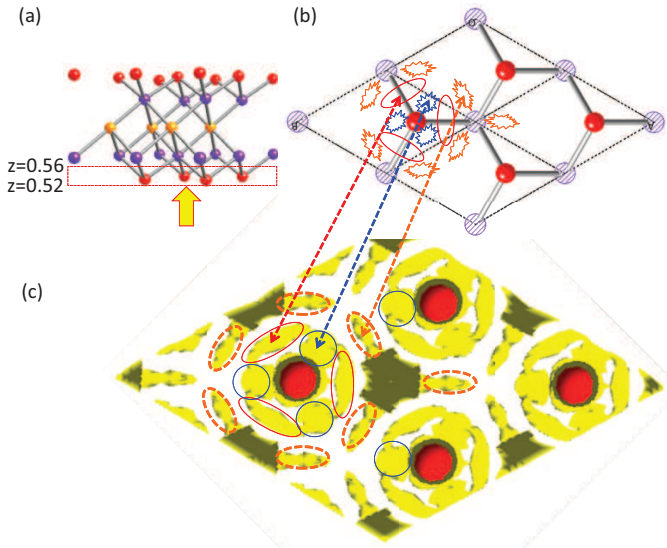}
\end{center}
\caption{\label{fig-Se1cut}(color online) The electron density contour map of Se1-layer, where (a) shows the side view of the ball and stick model of a quintuple layer cut between $z$=0.52 and 0.56, (b) shows the expected [001] projection of Se1 layer from below, with Bi atoms (dashed circles) behind the projection plane, and (c) is the corresponding integrated electron density contour map, where three $\sigma$-bonds (orange ovals) between Se1-Bi atoms and three dangling bonds (unpaired electrons) (in blue circle) per Se1 are indicated.  The electron clouds highlighted in orange color dashed ovals between Se1 atoms of six-fold symmetry are proposed coming from the two sets of $\pi$-bond electrons. }
\end{figure}

The crystal structure of implied atomic periodic arrangement has been determined with diffraction techniques with confidence, however, the actual bonding electron distribution existing between atoms remains unclear. In fact, X-ray incident beam of energy near $\sim$8 keV (copper target) is diffracted by the periodic electron density instead of the nuclei, i.e, the analysis based on Bragg law often ignores the periodicity of pure electron clouds, such as the electrons in the $\sigma$-bonds with their own periodicity as an ``electronic crystal". For the ideal covalent compound with atomic periodicity and pure $\sigma$-bond as an insulator, the  ``electronic crystal" shares identical symmetry to the ``atomic crystal", except for a small shift in real space depending on the degree of polarization  of the $\sigma$-bond being determined by the electron negativity.  For example, the Si crystal having perfect covalent bonds would have a $\frac{1}{2}$(Si-Si) distance shift between the ``electronic crystal" formed with $\sigma$-bond electron clouds relative to the ``atomic crystal", and there is no need to distinguish the two having identical point symmetry, especially when there is great difference on electron density.  In particular, the reciprocal $k$-space information obtained using diffraction technique does not distinguish the two contributions.  On the other hand, in the compound having mixed ionic and covalent bonds, or when a new type of chemical bond emerges, the ``electronic crystal" will not be necessary to have the identical point symmetry to the ``atomic crystal".  The most dramatic example can be found in the layered compounds having vdW gap between layers, such as the transition metal dichalcogenide (TMDC) of TiSe$_2$ and Bi$_2$Se$_3$, the valence electrons of Se across the vdW gap must not be able to form chemical bonds similar to those between intra-layer Bi-Se in the form of $\sigma$-bond.       

The real space electron density (ED) distribution can be extracted from the diffraction data with the aid of inverse Fourier transform of the X-ray diffraction data, i.e., converting the symmetry information from the reciprocal space in $\textbf{k}$ to the real space in $\textbf{r}$.  The theoretical foundation to apply Fourier transform analysis in crystallography has been documented thoroughly previously.\cite{EDtheory}  The bonding nature and the distribution of electrons condensed in the bonding region can be clearly visualized using this technique. The powerful ED mapping from the inverse Fourier transform has been demonstrated on exploring the nature of chemical bonds in Si crystal (covalent bond) and LiF (ionic bond),\cite{Takata1994} as well as in the layered material Na$_2$Ni$_2$TeO$_6$ and the prototype topological crystalline insulator Pb$_{1-x}$Sn$_x$Se.\cite{Shu2015, Karna2017}    

Beginning with the refined powder X-ray diffraction pattern measured at room temperature, the electron density distribution in real space can be extracted from the inverse Fourier transform of the X-ray diffraction data with the structure factors by employing the General Structure Analysis System (GSAS) program,\cite{Larson1990} as shown in Fig.~\ref{fig-EDin3D} for Bi$_2$Se$_3$.  The electron density contour map of a specific plane is obtainable by slicing the 3D electron density distribution below and above the selected atomic plane in the vicinity of $\pm$0.1 \AA.  The ED contour maps near the Se2 (z=0.62-0.68) and Se1 layers (z=0.52-0.56) are shown in Fig.~\ref{fig-Se2cut} and Fig.~\ref{fig-Se1cut}, respectively.  For the Se2-layer, the projection of electron cloud representing the paired electrons within six $\sigma$-bonds between Se2-Bi for each SeBi$_6$ octahedron can be identified clearly, which supports the validity of the proposed molecular model under the assumption of $sp^3d^2$ orbital hybridization.     

For the Se1-layer sitting near the vdW gap (see Fig.~\ref{fig-structure} and Fig.~\ref{fig-Se1cut}(a)), each Se1 atom facing the inner quintuple side is expected to form three $\sigma$-bonds with three Bi atoms, and the three unpaired electrons per Se1 on the surface side must contribute to the van der Waals attractive force bridging quintuple layers, or being dangling on the crystal surface. The three $\sigma$-bonds pointing toward the Bi atoms into the quintuple layer can be identified conclusively as the red ovals of three-fold symmetry, as compared in Fig.~\ref{fig-Se1cut}(b)-(c).  Since there are six unpaired electrons per Se1 ([Ar]3d$^{10}$4s$^2$4p$^4$), after three unpaired electrons per Se1 have formed three $\sigma$-bonds with the neighboring three Bi atoms in the quintuple layer, additional ED circled in blue color having a three-fold symmetry can be assigned corresponding to the remaining three unpaired electrons per Se1 reasonably.  While the unpaired electrons at the excited state are expected to be unstable and must seek for additional bonding in the condensed matter, electric dipole formation across the vdW gap is an option, otherwise the remaining unpaired electrons per Se1 must be exposed to the crystal surface as dangling bonds in the excited state.  

Following the assignment of three unpaired electrons in $sp^3d^2$ hybrid orbital to the $\sigma$-bond formation with three neighboring Bi atoms and three remaining unpaired for those  facing the vdW gap or the crystal surface, it is puzzling to find one extra set of ED having a six-fold symmetry centering at Se1 atoms (dashed orange ovals in Fig.~\ref{fig-Se1cut}(c)) can be identified, which is totally unexpected when all six valence electrons per Se1 are exhausted.  We speculate that the extra electron clouds of six-fold symmetry could be coming from an emergent chemical bond among surface Se1 atoms, i.e., a \underline{novel 2D electron condensation} may have occurred on the surface of Se1-layer facing the vdW gap side.  Judging from the possible side-to-side orbital overlap of the three half-filled $sp^3d^2$ lobes per Se1, it is possible that additional $\pi$-bonds are formed on both sides of each Bi$_2$Se$_3$ quintuple layer, similar to that proposed in graphene.\cite{Shu2015}  Since there are three unpaired electrons per Se1 on each side of a quintuple layer, a $\pi$-bond trimer of three-fold symmetry in 2D is expected.  However, in order to allow each $\pi$-bond trimer being shared by the six neighboring Se1 atoms in the same plane without breaking the six-fold atomic symmetry in 2D, it must be shared statistically as a conjugated $\pi$-bond trimer, which has been reflected on the six-fold symmetry of ED as shown in Fig.~\ref{fig-Se1cut}(c), instead of a three-fold symmetry expected for a static single $\pi$-bond trimer.  These results are consistent to the time lapse manner of X-ray diffraction data taking.  In addition, we postulate that a three-fold symmetry breaking could be observed with ultrafast probing techniques, which is supported indirectly by the calculated different bond lengths among carbon atoms via Monte Carlo simulation for graphene having a similar conjugated $\pi$-bond system.\cite{Fasolino2007}  Electron energy-loss spectroscopy (EELS) is used to explore the characteristics of the possible $\pi$-bond bonding electrons through the plasmonic absorption spectrum in the following.   


\section{\label{sec:level1} EELS studies of $\pi$ and $\sigma$ bonds in $Bi_2Se_3$  \protect\\}

\begin{figure*}
\includegraphics[width=5.0in]{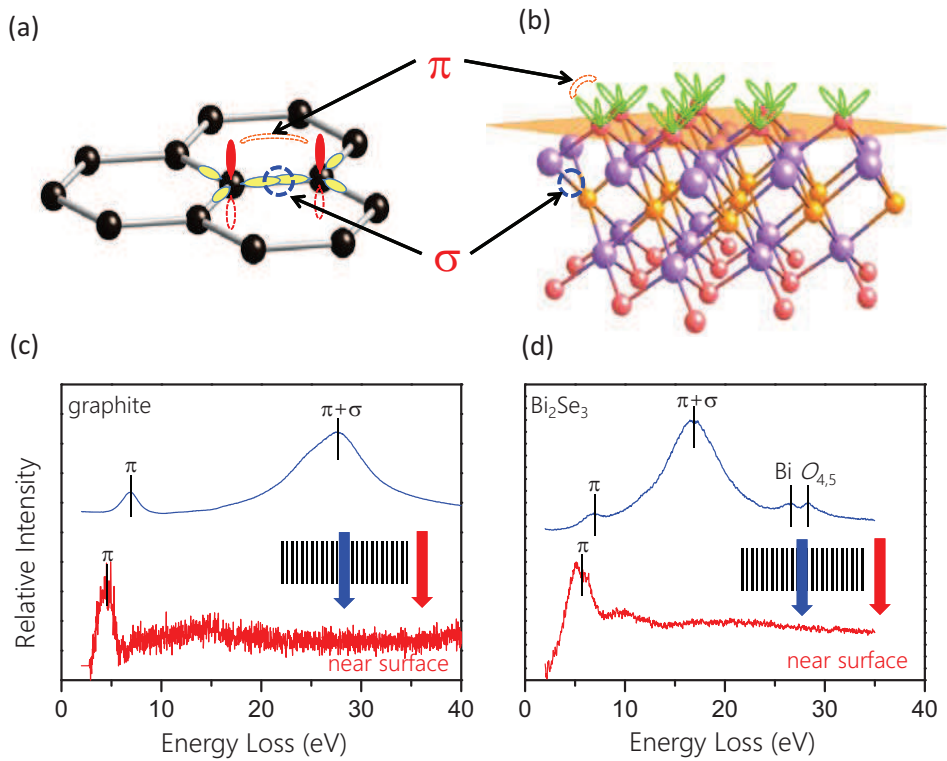}
\caption{\label{fig-EELS}(color online) The schematic drawings of $\pi$ and $\sigma$ bonds in real space for (a) graphene and (b) Bi$_2$Se$_3$.  The EELS spectra for (c) graphite and (d) Bi$_2$Se$_3$ are compared.  The EELS data for the surface-only excitation were taken $\sim$1 nm near the Se1 surface as shown schematically in the insets of (c)-(d), and the near surface energy loss reveals the excitation from the collective resonance of $\pi$ plasmons only.  }  
\vspace{-5mm}
\end{figure*}

Electron energy-loss spectroscopy (EELS) has been used to explore the kinetic energy loss of incident electron beam through energy absorption mechanisms, including intra- and inter-band transitions and excitations of phonon and plasmon origin.  In particular, the low-loss region (energy loss $<$ 50 eV) in the EELS spectra contains useful information concerning the collective modes of valence electron excitation, $e.g.$, surface and volume plasmons and the single-particle excitations like inter-band transitions and low-lying core-level ionizations.\cite{Raether1980, Egerton2011}  Although the plasmonic excitation of topological insulators has been explored with EELS and Terahertz experiments before,\cite{Politano2017, Viti2015, Politano2017a, Politano2015, Nechaev2015} we find that an intuitive understanding from the chemical bond perspective is lacking.  

The representative EELS spectra for graphite and Bi$_2$Se$_3$ are compared in Fig.~\ref{fig-EELS}(c)-(d).\cite{Liou2013, Carbone2009}  The spectral features near 7 and 27 eV for graphite have been demonstrated to be plasmon resonances related to the $\pi$ and ($\pi$+$\sigma$) electrons with effective number of electrons per atom of $n_\pi$ and ($n_\pi$+$n_\sigma$), respectively.\cite{Ichikawa1958, Taft1965}  The ratio of $n_\pi$:$(n_\pi$+$n_\sigma$) corresponding to the collective oscillation modes near 7 and 27 eV for graphite have been estimated to be $\sim$0.25 from the imaginary part of the EELS spectra, which is consistent to the expected value of $n_\pi$:$(n_\pi$+$n_\sigma$)=1:(1+3)=0.25 based on the proposed molecular model for graphite (graphene).\cite{Taft1965} The experimentally extracted $n_\pi$ and $n_\sigma$ values strongly suggest that the participating electrons for the assigned $\pi$ plasmon and the ($\pi$+$\sigma$) plasmon are coming from the unpaired electron in $p_z$ orbital and the three electrons in the hybridized $sp^2$ orbitals per carbon, respectively, as illustrated in Fig.~\ref{fig-EELS}(a).   

The EELS spectrum for Bi$_2$Se$_3$ shown in Fig.~\ref{fig-EELS}(d) reveals two main spectral features near 7 and 17 eV, which is in good agreement with those reported in the literature.\cite{Nascimento1999}   The two weak peaks at $\sim$26.4 and 28.4 eV can be identified as the Bi \textit{O}$_{4,5}$-edge excitation from Bi $5d$ electrons within an octahedral coordination.\cite{Liou2013, Nascimento1999}  Following the same interpretation that has been applied to the EELS spectrum of graphite (Fig.~\ref{fig-EELS}(c)) satisfactorily, we can tentatively assign the two main spectral features near 7 and 17 eV of Bi$_2$Se$_3$ to be volume plasmons of $\pi$ character and ($\pi$+$\sigma$) character, respectively.\cite{Liou2013, Huang2012}  

\begin{figure*}
\includegraphics[width=5.5in]{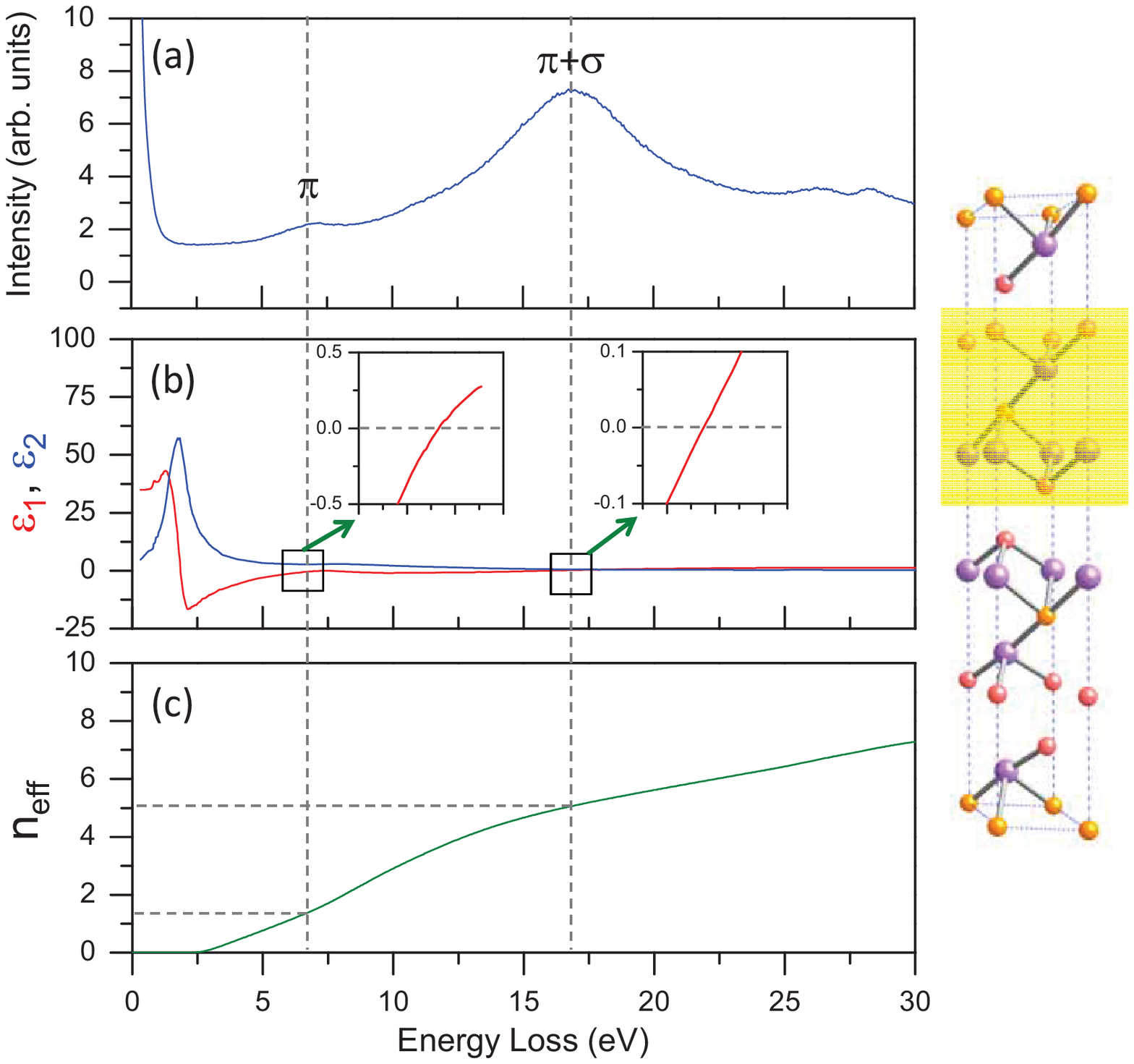}
\caption{\label{fig-EELS1}(color online) The raw data of EELS spectrum of Bi$_2$Se$_3$ is shown in (a), and the real ($\epsilon1$) and imaginary ($\epsilon2$) parts are shown in (b). The insets of (b) are blow-up views of $\epsilon1$ near zero, which suggest the existence of a collective oscillation (plasmon) when $\epsilon1$ passes zero with a positive slope.\cite{Raether1980}  For each 2D quintuple unit cell as highlighted in yellow on the right inset, there are two Bi and three Se atoms linked by 10 $\sigma$-bonds within and 6 unpaired electrons on the layer surfaces facing vdW gap without double counting, the effective number of electrons ($n_{eff}$) participating in the collective resonance are calculated from $\epsilon2$ (see Appendix) and shown in (c) to be $n_\pi$:$(n_\pi$+$n_\sigma$) $\sim$ 1.4 : 5.1.   }  
\vspace{-5mm}
\end{figure*}

Based on the proposed chemical bond model for Bi$_2$Se$_3$ having $\sigma$-bonds in the bulk and $\pi$-bonds on the surface facing vdW gap (Fig.~\ref{fig-structure}(b) and Fig.~\ref{fig-Se1cut}(b)), in order to count the effective number of electrons participating in the collective resonance of electrons in $\pi$- and $\sigma$-bonds per 2D quintuple unit (highlighted in the right inset of Fig.~\ref{fig-EELS1}), not the 3D unit cell that contains three sets of quintuples including two vdW gaps without $\sigma$-bonds in between (right inset of Fig.~\ref{fig-EELS1}), there would be 10 $\sigma$-bonds (20 paired electrons) between Bi-Se and 6 unpaired electrons for the two Se1 atoms per 2D quintuple unit without double counting. The expected ratio of $n_\pi$:$(n_\pi$+$n_\sigma$) would be 6:(6+20)=0.23 per 2D quintuple unit.  For the assigned collective oscillation modes near 7 eV ($\pi$ plasmons) and 17 eV ( ($\pi$+$\sigma$) plasmons) having effective participating numbers of $n_\pi$ and ($n_\pi+n_\sigma$), respectively, the ratio of $n_\pi$:$(n_\pi$+$n_\sigma$) has been estimated from the imaginary part of the EELS spectrum to be $\sim$ 1.4 : 5.1 = 0.27, as shown in Fig.~\ref{fig-EELS1}(c), which is in agreement with the predicted value within error.  In particular, when the vacancy of Se is counted,\cite{Huang2012} the ratio should be reduced and closer to the predicted ratio of 0.23 

Since $\pi$-bond has often been found and discussed in the systems like graphene and polymers for carbon atoms of particularly small atom size with partially filled $p$ orbital,\cite{Benzene, Fasolino2007} the finding of $\pi$-bond in a non-carbon system is rare, although evidence of $\pi$-bond existence has also been found in Pb$_{1-x}$Sn$_x$Se recently.\cite{Shu2015, Gunawan2014}  In order to confirm that $\pi$-bonds do exist near the surface, we have carried out scanning transmission electron microscopy (STEM) in an aloof geometry with the electron probe positioned at a distance of $\sim$1 nm from the surface, as illustrated in the inset of Fig.~\ref{fig-EELS}(c)-(d).\cite{Liou2013}  We find that the lower absorption peak representing $\pi$ plasmons is shifted to a slightly lower binding energy when EELS data were taken from the near-surface region for both graphite and Bi$_2$Se$_3$. In addition, the higher energy absorption that contains the mixing of $\sigma$ plasmons is significantly reduced, which implies that $\pi$-bonds must exist near the crystal surface only.  As a supporting evidence to the current STEM-EELS study results, we find that similar $\pi$- and $\sigma$-bonds from $sp^2$ and $sp^3$ hybridized orbitals have been identified by the time-resolved femtosecond EELS with convincing self-consistent density functional theory calculations.\cite{Carbone2009}  The latest momentum resolved inelastic electron scattering study also identified the acoustic plasmon mode on the surface of Bi$_2$Se$_3$. \cite{Jia2017}

\section{\label{sec:level1} Quantum chemistry calculation of surface $\pi$-bond\protect\\ }

\begin{figure}
\begin{center}
\includegraphics[width=3.5in]{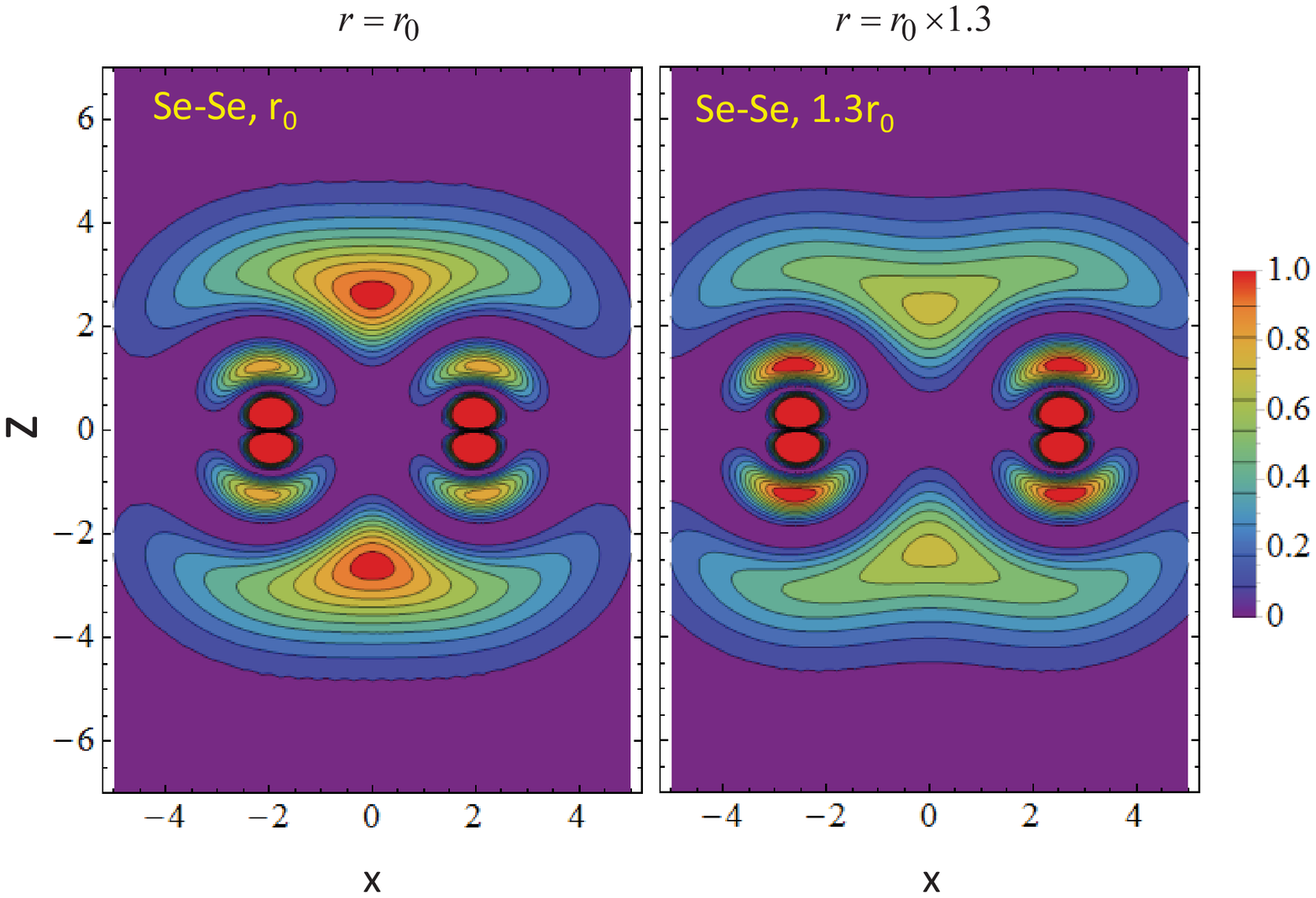}
\end{center}
\caption{\label{fig-4ppi}(color online) Starting from two half-filled $4p_z$ orbitals aligned along the $x$-direction, the calculated contours of the electron density distribution for two Se atoms separated by interatomic distances of $r_\circ$ and 1.3$r_\circ$($r_\circ$=3.93 \AA~ for Bi$_2$Se$_3$) are shown .      }
\end{figure}

The existence of $\pi$-bond in the topological crystalline insulator Pb$_{1-x}$Sn$_x$Se has been verified experimentally through electron density mapping and EELS.\cite{Shu2015}  The validity of $\pi$-bond formation via side-to-side half-filled 6$p_z$ orbital overlap at a critical distance between Pb-Pb was confirmed by the electron density contour  mapping through quantum chemistry calculation also.  As illustrated in Fig.~\ref{fig-structure}(b), current extended valence bond model for Bi$_2$Se$_3$ suggests that the Se1-layer near the vdW gap has a hybrid $sp^3d^2$ orbital per Se1. Three half-filled lobes of $sp^3d^2$ hybrid orbital per Se1 contribute to the three $\sigma$-bonds linking Se1-Bi within quintuple layer, and the remaining three half-filled lobes must point out facing the vdW gap or the crystal surface.  Similar to the 6$p_z$ orbital overlap as $\pi$-bond formation among Pb atoms in Pb$_{1-x}$Sn$_x$Se, we believe that the three unpaired electrons of hybrid $sp^3d^2$ orbital ($n$=4) on the surface of Se1-layer could also form $\pi$-bonds, which can be simplified and tested with the side-to-side 4$p_z$ orbital overlap of neighboring Se1 atoms in the same plane.  

For the surface Se-Se atoms in hexagonal close packing, the surface inter-Se1 distance is $\sim$3.93 \AA.   Quantum chemistry calculations for the surface Se-Se pair in Bi$_2$Se$_3$ have been performed using the 4$p_z$ wave functions.  The calculation started from the probability function ($|\Psi(x,y,z)|^2$) of the corresponding $n\pi$ bond orbital wave functions $\Psi_{n\pi}(x,y,z)$=$N(\psi_{n10}+\psi_{n 10}$), where $\psi_{n10}$ is an $np_z$ orbital function ($n$=4) for the atom of interest. The electron density distribution between $4p_z$ pairs as a function of inter-atomic distance is presented for Se-Se pairs in Fig.~\ref{fig-4ppi}.  The two atoms are located along the $x$-axis and the Slater rule\cite{Shu2015} is adopted for each atom to construct the $4p_z$ wave function.  It is apparent that $\pi$-bonds are formed by the two half-filled 4$p_z$ orbital pairs when the inter-Se distances are close to the equilibrium Se-Se bond lengths of $r_{\circ}$=3.93 \AA~for Bi$_2$Se$_3$, i.e., the outer shell electrons re-distribute from the localized near-nuclei region to outer shell region between Se-Se pair, which strongly supports the validity of the proposed $\pi$-bond formation on the surface of each Bi$_2$Se$_3$ quintuple layer.  In addition, the $\pi$-bond formation is found particularly sensitive to the interatomic distance for the Se-Se pair, as revealed by the significant drop of electron density near the proposed $\pi$-bond region when interatomic distance is enlarged by 10-30$\%$.  

  
\section{\label{sec:level1} real space view of symmetry-protected topological orders \protect\\ }

\begin{figure*}
\begin{center}
\includegraphics[width=5.5in]{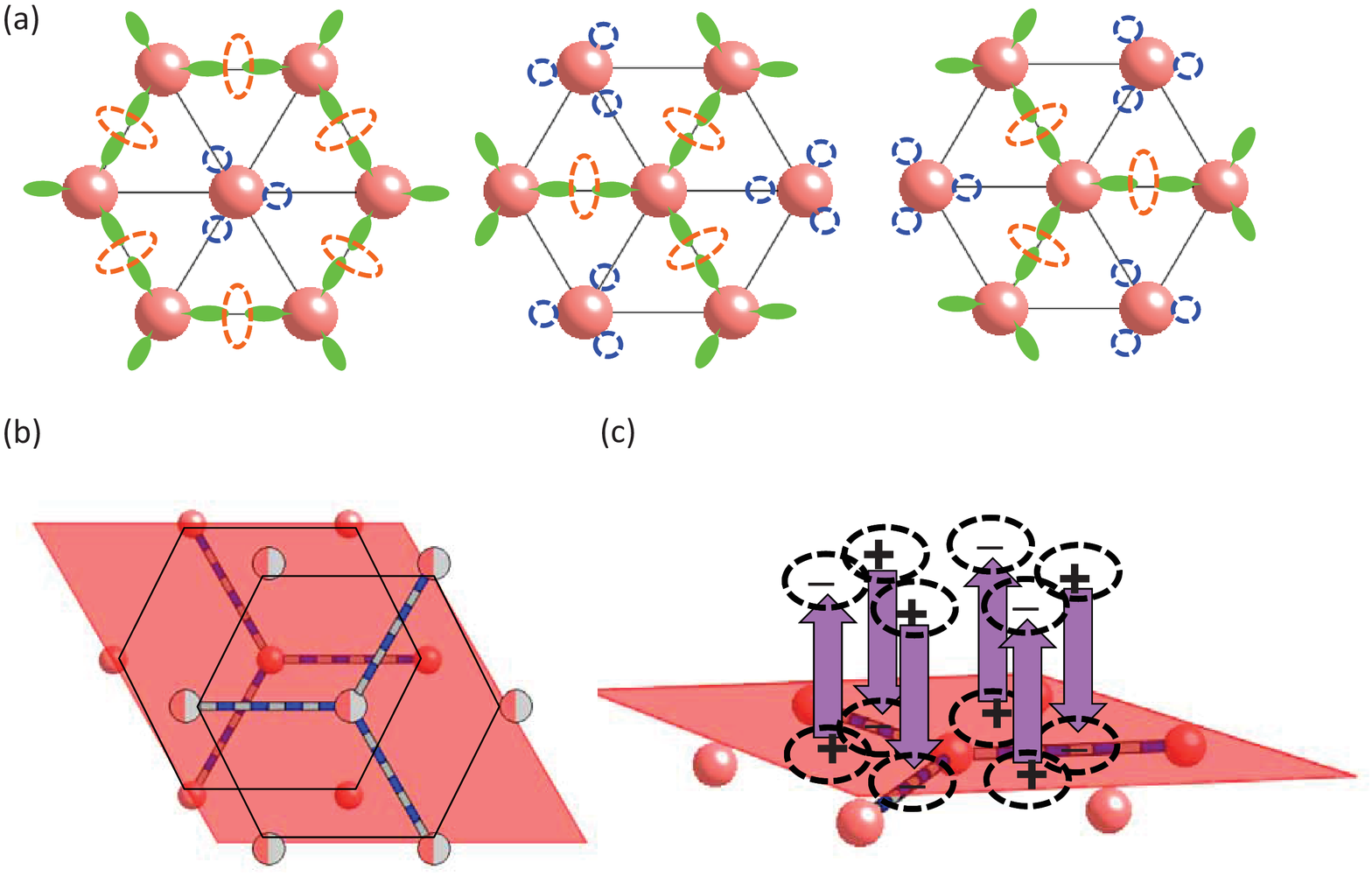}
\end{center}
\caption{\label{fig-conjugated1}(color online) (a) A top view of three possible arrangements of Se1 monoatomic layer.  Three lobes of half-filled $sp^3d^2$ hybrid orbital extend out of the basal plane (see Fig.~\ref{fig-structure}(b)), where $\pi$-bonds (orange dashed ovals) constructed by the three overlapped (green ovals) and unpaired (dashed blue circles) hybridized $sp^3d^2$ lobes per Se1 are proposed based on the ED contour mapping shown in Fig.~\ref{fig-Se1cut}. These three configurations are proposed to distribute randomly and entangle in space and time to form a conjugated $\pi$-bond trimer system with symmetry-protected topological orders.  (b) The top view of two adjacent Se1-layers across the vdW gap have $\pi$-bonds (dashed blue rods) arranged as a pair of trimers with minimized Coulomb repulsion. (c) The $\pi$-bond pairs form effective electric dipoles of anti-ferroelectric ordering across the vdW gap instantaneously, which is proposed to be the source of attractive vdW force to bind Bi$_2$Se$_3$ quintuple layers.  }
\end{figure*}

Based on the proposed valence bond model with added $\pi$-bond modification for Bi$_2$Se$_3$ (Fig.~\ref{fig-Se1cut} and Fig.~\ref{fig-EELS}), there are three unpaired electrons per Se1 facing the vdW gap or on the crystal surface.  However, there are six neighboring Se1 atoms in the same plane arranged in hexagonal shape without symmetry breaking, the proposed three $\pi$-bonds cannot satisfy all six Se1-Se1 pairs simultaneously.  If we choose to create a \underline{2D lattice of Se1 bonded with surface $\pi$-bonds only} and pick any Se1 as the origin, there would be three choices to arrange the unpaired electrons into the proposed surface $\pi$-bonds, i.e., the electrons condensed in $\pi$-bonds are allowed to fill the middle of all Se1-Se1 pairs within the 2D crystal unit kept in hexagonal symmetry, as shown in Fig.~\ref{fig-conjugated1}(a) with the dashed ovals.  It is important to note that although the three $\pi$-bond configurations for the Se1 hexagonal 2D unit seem to be different in real space locally, these three patterns can be generated via an adiabatic rotational or translational symmetry operation interchangeably.  In particular, these three configurations are degenerated at the same energy state when being extended to the infinity as a 2D condensed matter.  In an alternative description, the three real space electronic configurations via $\pi$-bond formation has a topological phase of multiple topological orders which are protected by the symmetry.  More specifically, Bi$_2$Se$_3$ has a topological phase with multiple topological orders protected by the time reversal symmetry, especially in view of the $\pi$-bond trimer under clockwise and anti-clockwise quantum fluctuation.  

In order for each $\pi$-bond trimer per Se1 being shared by the six neighboring Se1-Se1 pairs among the three possible configurations in real space, the $\pi$-bond trimer could form a conjugated system similar to that of the $\pi$-bond in graphene,\cite{Neto2009} except that the only one $\pi$-bond per carbon is shared by three C-C pairs in the honeycomb lattice in space and time statistically (see Fig.~\ref{fig-EELS}(a)), which allows $\pi$-bond to be formed randomly and dynamically, similar to the concept of a Resonating Valence Bond (RVB), or the three-electron chemical bond model discussed by L. Pauling in 1931.\cite{Pauling1931}  The $\pi$-bonds fluctuating in space and time can be viewed as a conjugated system with a symmetry-protected topological phase, similar to those observed in 1D conductive polymers.\cite{Shirakawa2003}  In fact, the 1D Majumda-Ghosh chain which contains spin singlet pairs can also be viewed as a conjugated system having translational symmetry-protected topological orders.\cite{Affleck1987}   Similar concept linking fractional charge and the topology in polyacetylene and graphene has been discussed before.\cite{Jackiw2007}  We propose that the surface $\pi$-bond trimer of Bi$_2$Se$_3$ could also form a conjugated system, i.e., even any one of the three real space configurations shown in Fig.~\ref{fig-conjugated1}(a) can be chosen as the ground state of the lowest enthalpy ($H$), especially in the occurrence of Peierls-like electron-phonon coupling that may lead to instantaneous 2D symmetry breaking from $C_6$ to $C_3$ symmetries upon electron condensation, the raised configurational entropy ($\Delta S$) can be used to lower the free energy further thermodynamically, as stated in the form of Gibbs free energy defined by $G=H-T\Delta S$.  

The ED contour mapping of Se1-layer cut (Fig.~\ref{fig-Se1cut}) revealed that all three configurations (Fig.~\ref{fig-conjugated1}(a)) are equally populated based on the preserved symmetries and intensities assigned to the $\sigma$-bonds in 3D and the $\pi$-bonds in 2D.  Since the observed ED mapping of Se1-layer is a cut from diffraction data taken within a time period, all symmetry information regardless possible different real space patterns are collected via inverse Fourier transform and being overlapped in the end.  For example, the blue circles in three-fold symmetry (see Fig.~\ref{fig-Se1cut} and Fig.~\ref{fig-conjugated1}) representing the distribution of electrons in the unpaired state, and the orange ovals of six-fold symmetry representing the two configurations of $\pi$-bond trimer of 60$^\circ$ rotation, which cannot be present simultaneously.  The observation that all three configurations are equally distributed statistically by the diffraction technique suggests a random distribution in space and time, in addition, all configurations must be long range entangled in space at any instance.  Interestingly, current real space chemical bond model containing surface conjugated $\pi$-bond system also implies the randomness on handedness for the $\pi$-bond trimer, which is equivalent to the theoretical description for a $Z_2$-type topological insulator containing a time-reversal symmetry protected topological orders.\cite{Hasan2010}

\section{\label{sec:level1} $\pi$-bond system on the crystal surface and in the vdW gap \protect\\ }

It is reasonable to assume that the conjugated $\pi$-bond system in the vdW gap may create an attractive force (vdW force) once effective electric dipoles are formed by the $\pi$-bond trimer systems across the vdW gap, as illustrated in Fig.~\ref{fig-conjugated1}(b)-(c).  The electric dipoles across the vdW gap are formed with the $\pi$-bond electron clouds in an anti-ferroelectric ordering to avoid Coulomb repulsion naturally.  In particular, the additional binding energy for the $\pi$-bond electrons in the vdW gap has been detected by the ARPES measurement to show the Dirac point sitting below the Fermi level,\cite{He2010, Hasan2010} which suggests that the bounded $\pi$-bond electrons in the vdW gap cannot be responsible for a conventional conduction as itinerant electrons described in the Fermi liquid theory. Instead, the unique surface conduction of Bi$_2$Se$_3$ could be coming from the conjugated $\pi$-bond system on the crystal surface mainly, which is similar to that of graphene as the prototype 2D conductor and the 1D conducting polymer.\cite{Roncali2007}   

Comparing the 2D contributions of $\pi$-bond electrons from the only one physical crystal surface and the multiple vdW gaps, the spectral weight of the former is expected to be weaker  in both ED and ARPES spectra. We speculate that 2D conduction for electrons bound in the vdW gaps remains possible, because the $\pi$-bond trimer pairs across the vdW gap are under constant quantum fluctuation within our proposed model, which allows local electron exchange in 2D persistently.  In addition, excitonic conduction is also possible for a narrow gap system with the help of SOC mechanism.  Similar argument can be applied to explain why the gapped bi-layer graphene and graphite are good conductors still.\cite{Zhang2009a}

It is proposed that the so-called surface conduction of a topological insulator could be misleading and should be viewed coming from a 2D conduction mechanism which is not necessarily on the crystal surface, especially when unpaired electrons of Se on the crystal surface can be easily bound with the reactive oxygen in the air in reality.  Above all, the linear dispersion of $E(\textbf{k})$ represents the energy-momentum relationship of quasi-free electrons confined in 2D plane, not necessarily on the physical surface of a 3D crystal, in contrast to the itinerant electrons of parabolic $E(\textbf{k})$ dispersion relationship found in a conventional 3D electronic system.  Both cases of gapless (i.e., cones touch at single Dirac point) or narrow gap with strong spin-orbit coupling (SOC) in 2D suggest an unique 2D conduction mechanism which is different from the itinerant electron concept applied to the conventional 3D system, the former has been identified in graphene successfully, and topological insulator could be a typical example for the latter. 

\section{\label{sec:level1} $Bi_2Se_3$ as a quasi-2D material \protect\\ }

\begin{figure*}
\begin{center}
\includegraphics[width=5.5in]{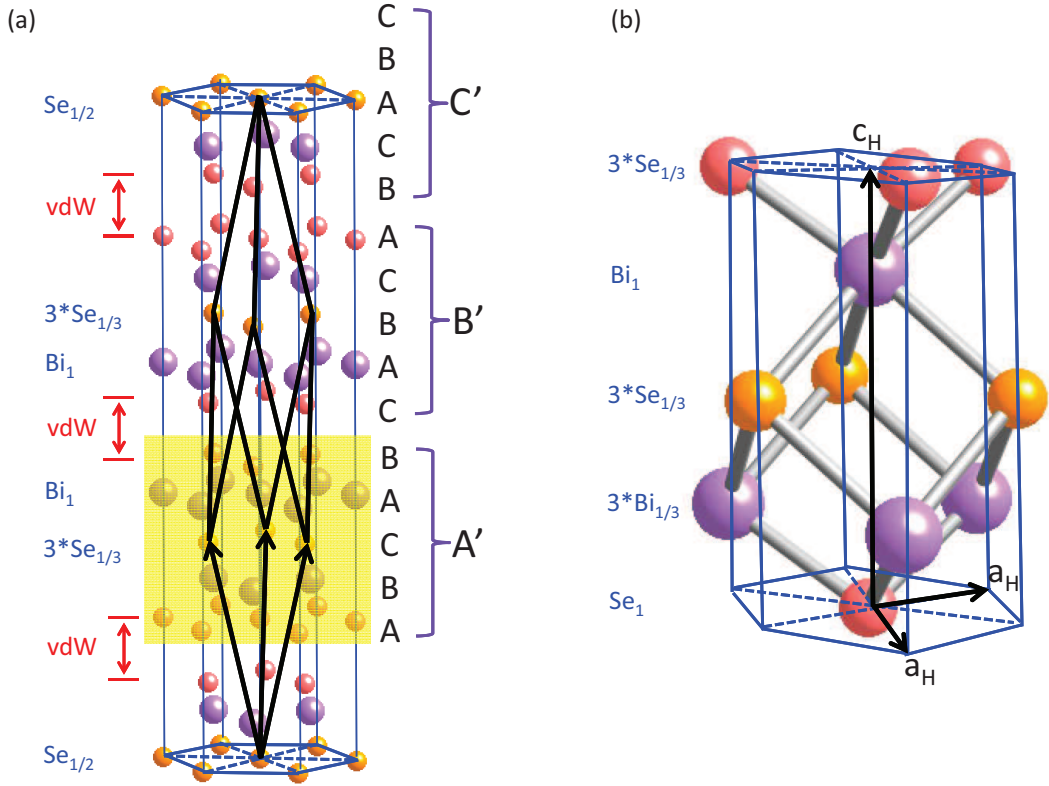}
\end{center}
\caption{\label{fig-Bi2Se3unitcell}(color online) (a) The 3D crystal unit cell of Bi$_2$Se$_3$ is described with space group $R\bar{3}m$ in hexagonal form of $a_H$=4.143~\AA~ and $c_H$=28.636~\AA, which contains three formula units per cell.\cite{Huang2012} The intra-layer stacking is shown in fcc close packing of ABCABC sequence and the quintuple layers are stacked in A$^\prime$B$^\prime$C$^\prime$A$^\prime$B$^\prime$C$^\prime$ sequence.  To complete a cycle of five monoatomic layers, the 3D primitive unit cell can be identified with the rhombohedral unit cell of $a_R$=9.840~\AA~ and $\gamma$=24.304$^\circ$, but it passes three vdW gaps without electron exchange.  (b) The 2D unit cell of Bi$_2$Se$_3$ is identified from the single quintuple layer (highlighted in (a)), which is allowed to extend along the $a_H$-direction but must terminate at one $c_H$ without extension and sharing. }
\end{figure*}

 The 3D atomic crystal structure of Bi$_2$Se$_3$ has been described with the space group $R\bar{3}m$ through X-ray and neutron diffraction experiments convincingly.  Since there are three formula units per 3D unit cell as shown in  Fig.~\ref{fig-Bi2Se3unitcell}(a), a primitive rhombohedral unit cell has been chosen for the Brillouin zone in $k$-space constructed for band picture calculation.\cite{Zhang2009}  While it is common to derive the band structure of a crystalline material following its real space crystal symmetry, i.e., using the atomic planes of $d_{hkl}$-spacing with corresponding normal vector in $\textbf{k}$-direction, it is under the assumption that band structure can be constructed via infinite extension of unit cells from discrete energy levels to band continuum along three dimensions in real space.  However, there is no inter-layer electron exchange for 2D materials across the vdW gap, i.e., the Coulomb-type interaction is an ``action at a distance" between layers, which implies that the band calculation based on the assumption of infinite extension along the inter-layer $c$-direction should be re-examined.  
 
Strictly speaking, Bi$_2$Se$_3$ should be categorized as a quasi-2D crystalline material of finite penta-atomic layer thickness, as shown in Fig.~\ref{fig-Bi2Se3unitcell}, being consistent with the ideal 2D material graphene of monoatomic layer.  The fcc (ABCABC) close packing of five Bi and Se atomic planes form a quintuple layer, and all quintuple layers show inter-layer A$^\prime$B$^\prime$C$^\prime$A$^\prime$B$^\prime$C$^\prime$ stacking as well to close the quintuple cycle of five (2$\times$Bi+3$\times$Se) monoatomic planes, which leads to three quintuple layers per 3D unit cell.  It should be noted that the A$^\prime$B$^\prime$C$^\prime$A$^\prime$B$^\prime$C$^\prime$ inter-layer stacking has a Coulomb type interaction between Se-layers across the vdW gap, except for the dielectric breakdown under ultrahigh electric field or the occurrence of excitonic conduction under high temperature and narrow gap conditions, there should be no electron exchange or wave function overlap exists between quintuple layers in normal conditions.  Although the atomic ordering along three dimensions follows the point symmetry operation of $R\bar{3}m$ exactly, there is no electronic correlation across the three vdW gaps per 3D unit cell, as depicted by the chosen rhombohedral primitive unit cell (Fig.~\ref{fig-Bi2Se3unitcell}(a)).  Since the electron exchange is restricted within each quintuple layer, infinite extension along the $a_H$-direction is allowed, but the $c_H$-direction must be terminated to one penta-atomic thickness only, as illustrated in Fig.~\ref{fig-Bi2Se3unitcell}(b) by the 2D unit cell of Bi$_2$Se$_3$.  As a result, all quintuple layers are independent and electronically equivalent, the A$^\prime$B$^\prime$C$^\prime$A$^\prime$B$^\prime$C$^\prime$ inter-layer stacking has no electronic correlation at all but a Coulomb-type interaction of action at a distance.  

Based on the analysis shown above, the ARPES observed band picture should be interpreted coming from the energy dispersion relation $E$($\textbf{k})$ of \underline{bound electrons per quintuple layer with finite thickness}, instead of coming from a crystal of implied 3D crystal symmetry having assumed electron correlation between layers.  The key difference between the conventional 3D material and the quasi-2D material like Bi$_2$Se$_3$ is the finite thickness for the latter.  Although it is common to apply Fourier transform technique using the real space 3D crystal symmetry to generate the corresponding BZ shown in the reciprocal space, it could be misleading when applying the same procedure to the quasi-2D material of finite thickness per layer.  We propose that a new band calculation for Bi$_2$Se$_3$ is required, which must be based on a proper choice of 2D primitive unit cell from the single quintuple layer of finite penta-atomic thickness, instead of using the 3D rhombohedral primitive unit cell that passes through three vdW gaps of vacuum without electron exchange.  

\section{\label{sec:level1}Conclusions\protect\\}

Based on the integrated study of crystal symmetry, surface atomic coordination, and outer-shell valence electron distribution, we have proposed an extended valence bond model to interpret the unique physical properties of  Bi$_2$Se$_3$ being an important candidate of high $ZT$ thermoelectric material and also a $Z_2$-type topological insulator.  Conjugated $\pi$-bond system on the surface of each Bi$_2$Se$_3$ quintuple layer has been proposed to be responsible for the 2D conduction mechanism similar to that of graphene.  Supporting experimental evidences for the existence of a 2D conjugated $\pi$-bond system in the vdW gap and on the crystal surface are provided by the volume plasmons detected with EELS studies and the electron density contour mapping obtained from the inverse Fourier transform of X-ray diffraction.  The proposed extended valence bond model has provided a powerful real space interpretation to the Dirac cone energy dispersion of a 2D electron system.  In addition, the proposed $\pi$-bond system is able to provide a reasonable explanation to the origin of vdW force for the layered compounds for the first time.  We believe the inclusion of the concept of conjugated $\pi$-bond system in real space view is valuable to understand why topological material is important, and how to develop more for application from the material science and chemical perspectives. 

\section{Appendix: energy loss due to plasmon excitation}

Plasmon can be understood as collective longitudinal charge-density oscillation of free electron gas in the bulk induced by the passing incident electron beam. The free electrons in the solids are initially repelled and displaced from their original positions, but the displaced electrons will attempt to restore to their original positions after the bombardment of the incident electron beam.  The resulting oscillation process in the form of local electron density variation has a characteristic frequency to be quantitatively described by the dielectric theory. 

When a beam of incident electrons of energy $E_\circ$ and momentum $\hbar k_\circ$ penetrates the material, scattering by an inelastic event results in a loss of energy $\hbar \omega$ ($\Delta E$) and momentum transfer $\hbar q$. The differential inelastic scattering cross-section of a material can be characterized by a complex dielectric function $\epsilon$($q$, $\omega$) as\cite{Raether1980}
\begin{equation}
\frac{d^2\sigma}{d\omega d\Omega}\approx\frac{1}{q^2}Im[\frac{-1}{\epsilon(q, \omega)}],
\end{equation}  
\noindent where $\epsilon$($q$, $\omega$)=$\epsilon_1$($q$, $\omega$)+$i$$\epsilon_2$($q$, $\omega$) and $Im[\frac{-1}{\epsilon(q, \omega)}$] = $\frac{\epsilon_2}{\epsilon_1^2+\epsilon_2^2}$ is called the energy-loss function. The energy loss function appears as a weak $\epsilon2$ peak at the plasma frequency ($\omega_p$) and the real part of the dielectric function $\epsilon1$($\omega_p$)=0 has a positive slope.  The effective number of electrons per formula unit participating in the plasmonic oscillation ($n_{eff}$) can be estimated via 
\begin{equation}
n_{eff} = \frac{m_\circ}{2\pi^2e^2N}\int_0^\omega \omega^\prime \cdot Im[\frac{-1}{\epsilon(q, \omega^\prime)}] d\omega^\prime, 
\end{equation}  
\noindent where $m_\circ$ is the free electron mass and $N$ is the density of atoms in the material, and the effective number of electrons contribute up to the frequency $\omega$.

\section*{Acknowledgments}
FCC acknowledges the support provided by the Ministry of Science and Technology in Taiwan under project number MOST-102-2119-M-002 -004.  GJS acknowledges the support provided by MOST-Taiwan under project number 103-2811-M-002 -001.


\pagebreak

\end{document}